\documentclass[twocolumn,secnumarabic,amssymb, nobibnotes, aps, prl, superscriptaddress]{revtex4-1}

\usepackage{graphicx}
\usepackage{color}

\usepackage{lineno}
\begin{document}

\title{Spectral coherence properties of continuum generation in bulk crystals}

\author{Benjamin Maingot}
\email{Benjamin.Maingot@fastlite.com}
\affiliation{Institut de Physique de Nice (INPHYNI), Universit\'e C\^{o}te d'Azur, CNRS, UMR 7010, 1361 route des Lucioles, 06560 Valbonne, France}
\affiliation{Fastlite, 165 route des Cistes, 06600 Antibes, France}

\author{Gilles Ch\'eriaux}
\affiliation{Institut de Physique de Nice (INPHYNI), Universit\'e C\^{o}te d'Azur, CNRS, UMR 7010, 1361 route des Lucioles, 06560 Valbonne, France}

\author{Nicolas Forget}
\affiliation{Fastlite, 165 route des Cistes, 06600 Antibes, France}

\author{Aur\'elie Jullien}
\email{Aurelie.Jullien@inphyni.cnrs.fr}
\affiliation{Institut de Physique de Nice (INPHYNI), Universit\'e C\^{o}te d'Azur, CNRS, UMR 7010, 1361 route des Lucioles, 06560 Valbonne, France}

\begin{abstract}
The stability of the phase difference between two white-light continua, generated from the same 180-fs pulses at $\simeq$1035\,nm, is assessed by a modified Bellini-Hansch interferometer. Mutual spectral stability is studied as a function of several parameters: position of the nonlinear crystal with respect to the beam waist, interaction length, and pulse energy. Intensity-to-phase coupling coefficients are measured and stability regions are identified. 
\end{abstract}

\maketitle

Third-generation femtosecond sources \cite{Fattahi2014} rely on white-light continuum generation (WLG) to extend the spectrum of (sub) picosecond optical pulses at 1030 nm. By focusing $\mu$J-level pulses in a nonlinear transparent crystal such as YAG, a broad and stable spectrum extending from the visible to 1.5$\mu$m can be sustained \cite{bradler2009}. Among the essential properties of WLG is the pulse-to-pulse phase relationship between the spectral components of the continuum, which may be referred to as the intrapulse coherence \cite{Raabe2017}. Intrapulse coherence plays an essential role not only in pulse compression but also in the generation of pulses with a stable phase relationship between the optical carrier and the pulse envelope (carrier-enveloppe phase or CEP). Indeed, as described in \cite{Baltuska2002}, the frequency difference (DFG) between two waves sharing common phase fluctuations is intrinsically CEP-stable \cite{Baltuska2002,Cerullo2011}. In most CEP-stable sources, these two waves are either two parts of the continuum, or the driving pulse at 1030 nm and the visible/infrared wing of a continuum excited by the very same pulse. In both cases, intrapulse decoherence could be a source of CEP noise in the DFG process and/or be an additional source of noise in the f-to-2f interferometers used to measure the CEP drift. Even if long-term CEP noise as low as 65 mrad rms \cite{Thire2018} has been demonstrated, the origin and level of the CEP noise floor remains an open question. As pointed out by numerical simulations, the plasma contribution in filaments could be a major source of loss of intrapulse coherence \cite{Raabe2017}.

In this paper, we experimentally characterize the intrapulse coherence of a continuum generated in a thick YAG crystal by an amplified Ytterbium laser. The characterization method relies on a variant of the Bellini-Hansch interferometer \cite{bellini2000} : two identical WLGs are inserted in a balanced Mach-Zhender interferometer and the shot-to-shot phase fluctuations are analyzed by spectral interferometry. The relative phase noise is frequency-resolved and the effects of seed energy, crystal position, and crystal length are assessed. From this multi-parameter study we define stability ranges for WLG, assess intensity-to-phase coupling coefficients \cite{Baltuska2003}, and identify an operating range in which these coefficients vanish.

\begin{figure}[htp]
\includegraphics[width=1\linewidth]{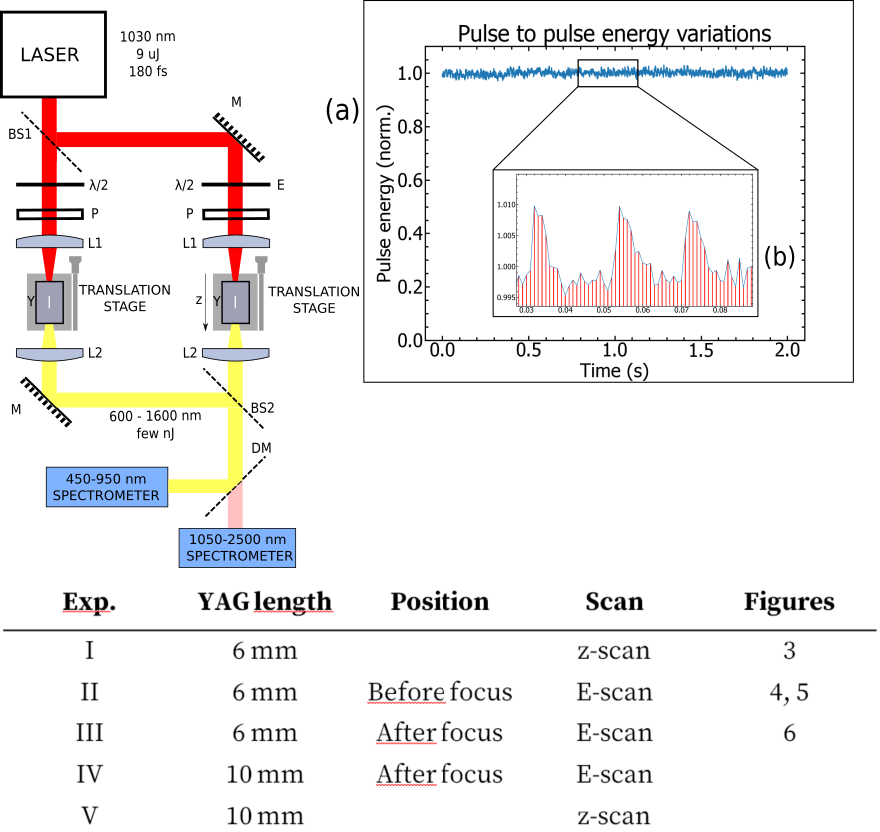}
	\caption{Diagram of the interferometer. M: mirrors, BS1: 1030 nm 50/50 beam splitter, BS2: low GDD 600-1500 nm 50/50 beam splitter, P: polarizer, L1: focusing lens 150 mm, L2: achromat focusing lens  50 mm, Y: YAG crystal, DM: dichroic mirror. (a) Pulse-to-pulse energy stability of the driving laser. (b) Zoom over 100 ms (each red bar is a laser shot). The table summarizes and labels the different experimental configurations and parameter scans performed in this study. }
	\label{diagram}
\end{figure}

The generation parameters for WLG follows the general guidelines indicated by \cite{Calendron2015} for sub-ps WLG. We use Yttrium Aluminum Garnet (YAG) crystals and the seed laser is focused with a numerical aperture of 0.015 to optimize the generation of the infrared part of the spectrum. Two configurations are distinguished in this work, depending on the position of the crystal relatively to the laser focus. When the front face of the crystal is placed before the waist, the broadening is limited in the infrared and the visible wing of the continuum is broad and stable. When placed after the waist, the infrared wing is extended and the visible wing tends to become unstable. In addition, two crystal lengths are studied: 6 mm and 10 mm. Increasing the crystal length tends to increase the intensity and the width of the infrared spectrum.

The pump laser is a regenerative CPA system (Pharos, Light Conversion) delivering pulses at $\simeq$1035\,nm with a pulse duration of $\simeq$180\,fs FWHM. The beam diameter (1/$e^2$) is 4.5\,mm with excellent beam quality (M$^2$=1.1). For these experiments the repetition rate is set at 1 kHz and $\simeq$9\,$\mu$J are sampled from the available $\simeq$1\,mJ of pulse energy. The interferometer is of Mach-Zendher type. Each arm includes a half-waveplate and a polarizer to adjust the pulse energy on each YAG crystal independently. A pair of f=150\,mm lenses focus the beam in identical 6-mm-long YAG crystals (NA: 0.015) and f=50 mm achromatic lenses collimate the spectrally-broadened beams. One arm of the interferometer serves as a test arm whereas the other one is the reference arm (Fig. \ref{diagram}). The adjustable parameters in the test arm are reported in the table of Fig. \ref{diagram}. Although not shown in Fig. \ref{diagram}, two reflective delay lines control the relative group delay of the two arms.
%Furthermore, the energy is set below the double filamentation threshold as this gives the broadest spectrum and is within the window of phase stability CITE BALTUSKA. 
Once recombined, the beams are focused into two array-based spectrometers: a 450-1100\,nm Silicon spectrometer (resolution of 0.26 nm, integration time of 1.1 ms) and a 1000-2500 nm InGaAs spectrometer (resolution of 3.27 nm, integration time of 1 ms). These spectral ranges will be referred to as the short wavelength range and the long wavelength range.
%A 950\,nm short-pass dichroic  mirror and a silicon plate filter out the pump.

For a relative group delay of a few hundreds of fs, the measured spectra show interference fringes over the full spectral range. The spectral phase is extracted from each single-shot interferograph by a discrete Hilbert transform \cite{Lepetit1995,Borzsonyi2013}. To study the relative phase stability, 1000 consecutive single-shot spectra are acquired and analyzed. The standard deviation of the spectral phase at each wavelength is used as metric to assess the intrapulse coherence.
Prior to the computation of the standard deviation, a 5-Hz high-pass filter is applied to remove slow phase drifts. The rationale for this choice is that (i) most, if not all, CEP-stable sources include a slow feedback loop \cite{Forget2009}, (ii) we focus on shot-to-shot phase fluctuations.
Wavelengths for which the spectrometer noise equals or exceeds the detected level of light were discarded from this analysis.

%\begin{figure}[htp]
%\includegraphics[width=1\linewidth]{Fig2.png}
%	\caption{(a,c) Typical spectrograms over 1000 single shot measurements, in the short (a) and long (c) wavelength range.  (b,d) Standard deviation of the phase retrieved from the spectrograms in (a, c) as a function of wavelength with a 5 Hz high pass filter alone (1) and with an additional filtering on the fast energy peaks of the pump laser (2). The absorption window of silicon is also highlighted to discard the measurements in this range because of the low signal to noise.}
%	\label{spectrograms}
%\end{figure}

The shot-to-shot energy stability of the Pharos system is shown in Figure \ref{diagram}. Despite an overall stability below 0.5\% rms, 1\% peak-to-peak fluctuations followed by a fast relaxation (i.e. over few laser shots) are observed. These events appear randomly and this behaviour is not explained at this time. At first, these pulses are discarded for the stability analysis but then used to assess the intensity-to-phase transfer coefficient \cite{Baltuska2003}.

To subtract the phase artifacts from the optics and assess the precision of the measurement, we proceed in two steps: (i) without YAG crystals and (ii) with YAG crystals and balanced WLGs.
Without crystals the  typical measured standard deviation is <5\,mrad at 1030\,nm, which provides a ground floor for the phase noise measurement.
For (ii), the front face of the 6-mm YAG crystal is placed slightly before the waist with the energy set slightly below the double filamentation threshold. 
This configuration, which corresponds to the stable range of \cite{Baltuska2003} is chosen as the operating point of the reference arm in the following. 
Figure \ref{spectrograms}a and \ref{spectrograms}c show the recorded spectrograms. Figures \ref{spectrograms}b and \ref{spectrograms}d show the respective phase stability with and without removing the 1\% energy overshoots.
Compared to the ground floor, the phase noise is increased to <20\,mrad and the phase stability tends to decrease as the wavelength offset from the pump wavelength increases.

\begin{figure}[htp]
\centering{\includegraphics[width=1\linewidth]{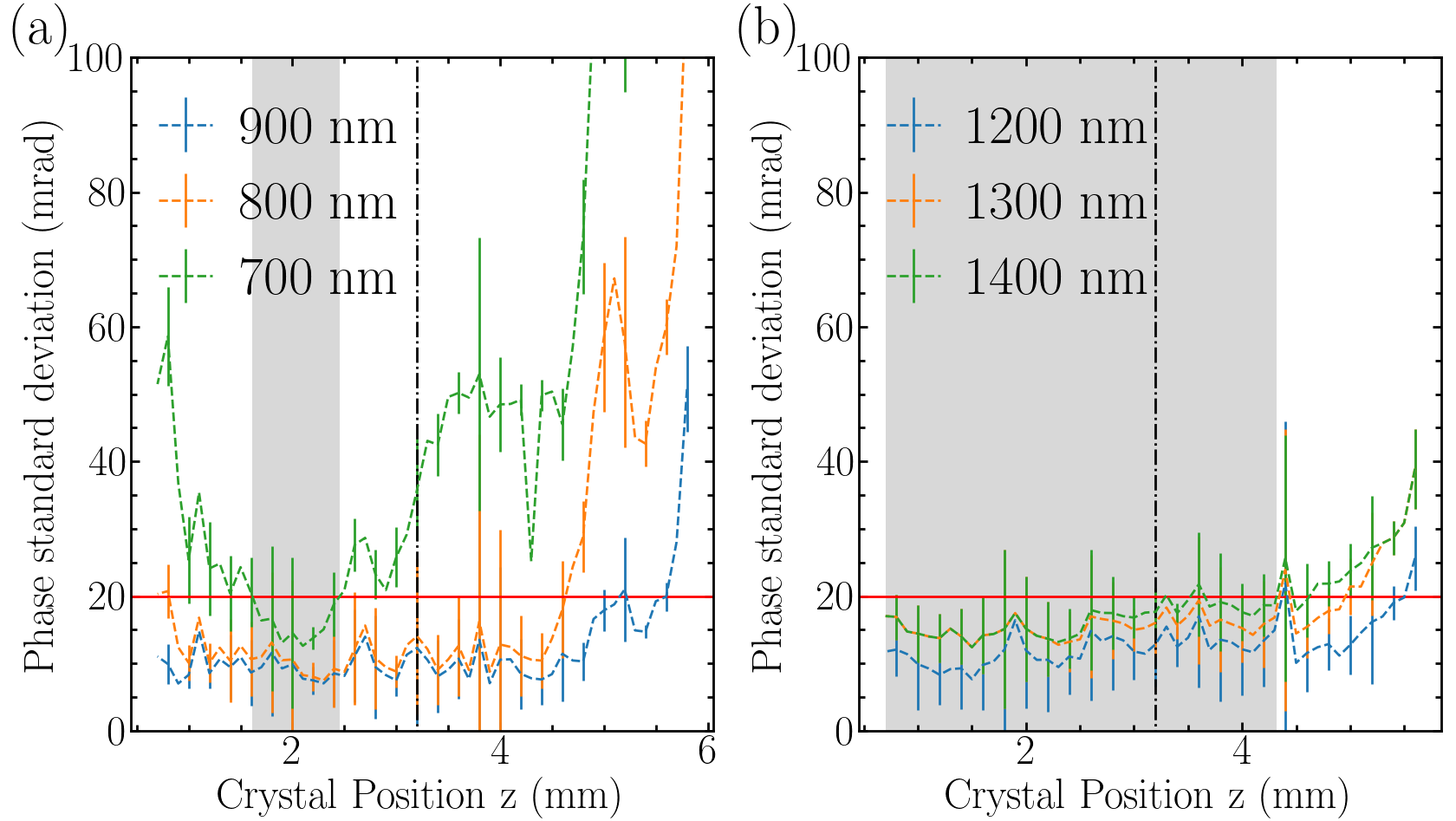}}
	\caption{(I) Phase standard deviation as a function of crystal position in the short (a) and long (b)  wavelength range. The focus (z=3.2 mm) is indicated with a dashed line.  Lower z-values values gets the YAG closer to the focusing lens while higher z-values are further away from the lens. The red line indicates the balanced stability of the interferometer. The grey areas refer to the stability operating zone. }
	\label{position}
\end{figure}

The first investigated parameter is the position of the YAG crystal along the propagation axis (I). Figure \ref{position} shows the phase stability as a fonction of the crystal position for three selected wavelengths in the short wavelength range (Fig. \ref{position}a) and in the long wavelength range (Fig. \ref{position}b). 
The long wavelength side shows a wide stability range of 4 mm (from $z=$ 0.2 to $z=$ 4.2)  with a phase noise below 20 mrad rms. Although the crystal position modifies the spectral shape, almost no influence on the phase noise is measured, except when the crystal is moved far from the waist.
Conversely, the stability range for the short wavelength range is much narrower (1 mm around $z=$ 2\,mm). This range coincides  with the generation of an intense, broad and stable spectrum in the visible. Out of this range, the phase noise tend to increase (at 700\,nm especially), even if a second noise plateau can be identified at $z=$ 4\,mm.

To study the influence of pulse energy, both crystals are set to $z=$ 2\,mm (II) and then to $z=$ 4\,mm (III). 
For $z=$ 2\,mm (II), the energy is varied from 0.35 to 1.57\,$\mu$J (step of 20\,nJ steps), that is, from the onset of filamentation to above the threshold of double filamentation (1.13 $\mu$J). Double filamentation is easily identified via the deep modulation of the spectrum caused by the interference between the two filaments. 
Both short and large wavelength ranges display the same qualitative behavior: the phase noise first decreases and then increases with pulse energy. As identified in \cite{Baltuska2003}, the phase noise is low, 20 mrad rms for the visible range and 30 mrad rms for the IR range, when the single filament is stable. The stability range extends from 0.66 to 1.15 $\mu$J in the short wavelength range and from 0.83 to 1.11 $\mu$J in the long wavelength range. In both cases, the stability range extends up to the threshold of double filamentation.  

\begin{figure}[htp]
\centering{\includegraphics[width=1\linewidth]{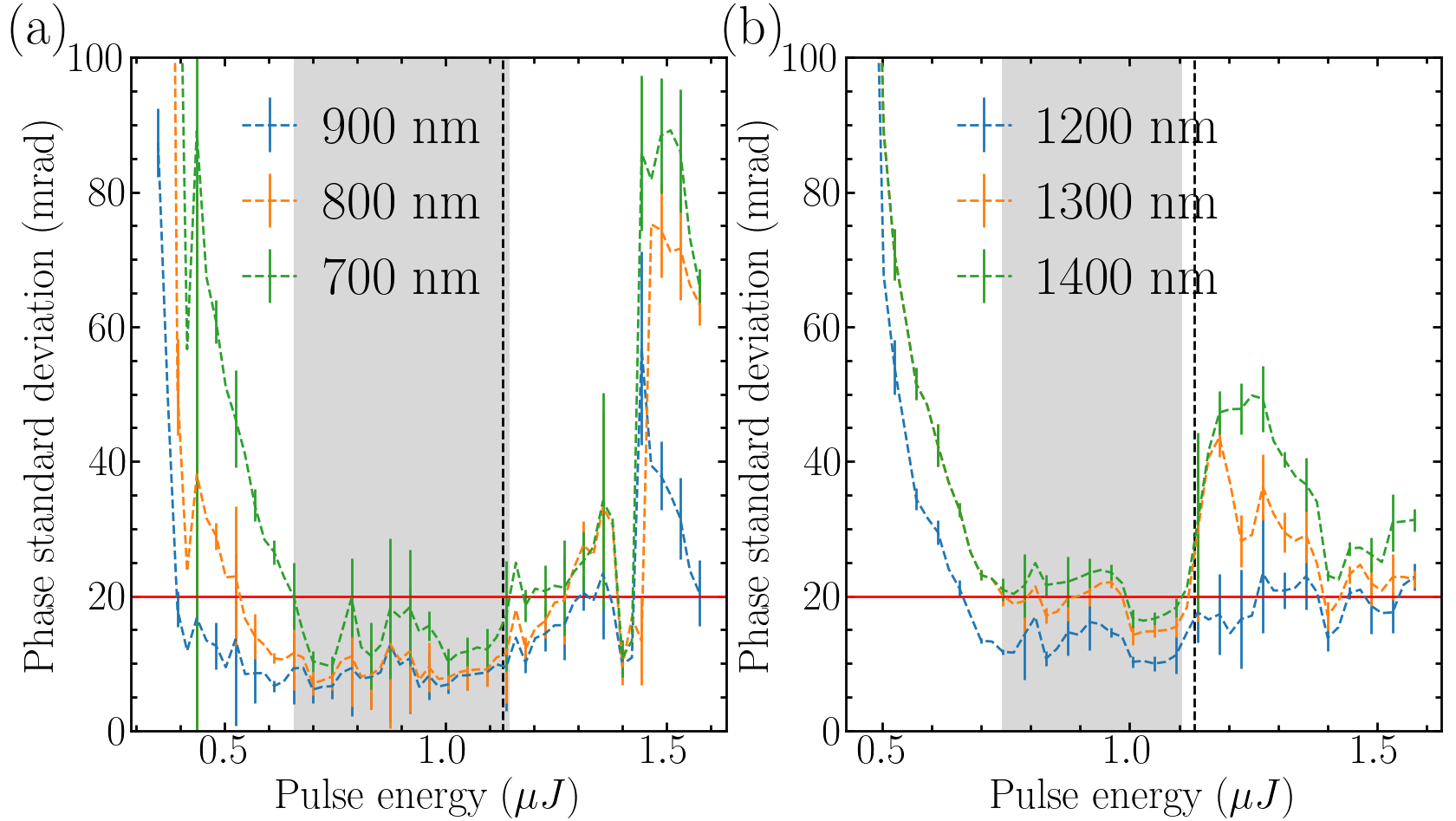}}
	\caption{(II) Phase standard deviation as a function of seed pulse energy in the short (a) and long (b) wavelength ranges. The energy axis starts at the onset of filamentation. The dotted black line refers to the onset of double-filamentation. The grey areas refer to the stability operating zone.}
	\label{before}
\end{figure}

For $z=$ 4\,mm (III), which is the position for which the infrared wing is favoured, the energy was varied from 0.5 to 1.23\,$\mu$J (15\,nJ steps). The results are very similar: noise decreases with increasing energy and reaches a stability plateau until double filamentation breaks in. However, as expected from Fig. \ref{position}a, by comparison with configuration (II), the range where the noise is below 20 mrad is narrower in the short wavelength range (0.89 - 1.1$\mu$J) and slightly larger in the long wavelength range (0.79 - 1.1 $\mu$J). 

The last parameter studied in this paper is the crystal length. Here we compare a 6\,mm crystal with a 10\,mm crystal, all other parameters being equal. With a longer crystal, the infrared spectrum is broader and the infrared cut-off wavelength (measured at 20\% of the maximum spectral density) increases from 1580\,nm to 1720\,nm. The behaviours are qualitatively similar. Quantitative results are summarized in Table \ref{summary}.

\begin{table*}[htp]
	\begin{tabular}{|c|ccccc|}
		\hline
		& I  & II  & III & IV & V \\  
		\hline\hline
		Short wavelengths range phase stability & 1.6 - 2.5 mm & 0.66 - 1.15 $\mu$J & 0.89 - 1.1 $\mu$J & above 20 mrad & above 20 mrad\\
		\hline
		Long wavelengths range phase stability & 0.7 - 4.32 mm & 0.83 - 1.11 $\mu$J & 0.79 - 1.1 $\mu$J & 0.83 - 1.3 $\mu$J & 0.5 - 4.8 mm\\
		\hline
		min jitter @750 nm & 12 mrad & 9 mrad & 9.5 mrad & 28 mrad & 21 mrad\\
		\hline
		min jitter @1300 nm & 12 mrad & 14 mrad & 12 mrad & 5 mrad & 7 mrad\\
		\hline
		$\kappa$ @750 nm, middle of stab. range & 100 mrad/\% & 130 mrad/\% & 15 mrad/\% & 25 mrad/\% & 64 mrad/\%\\ 
		\hline
		$\kappa$ @1300 nm, middle of stab. range & below noise & below noise & 27 mrad/\% & 5 mrad/\% & 95 mrad/\%\\ 
		\hline
	\end{tabular}
	\caption{Significant phase stability values for the different experimental configurations.}
	\label{summary}
\end{table*}

\begin{figure}[htp]
\includegraphics[width=1\columnwidth]{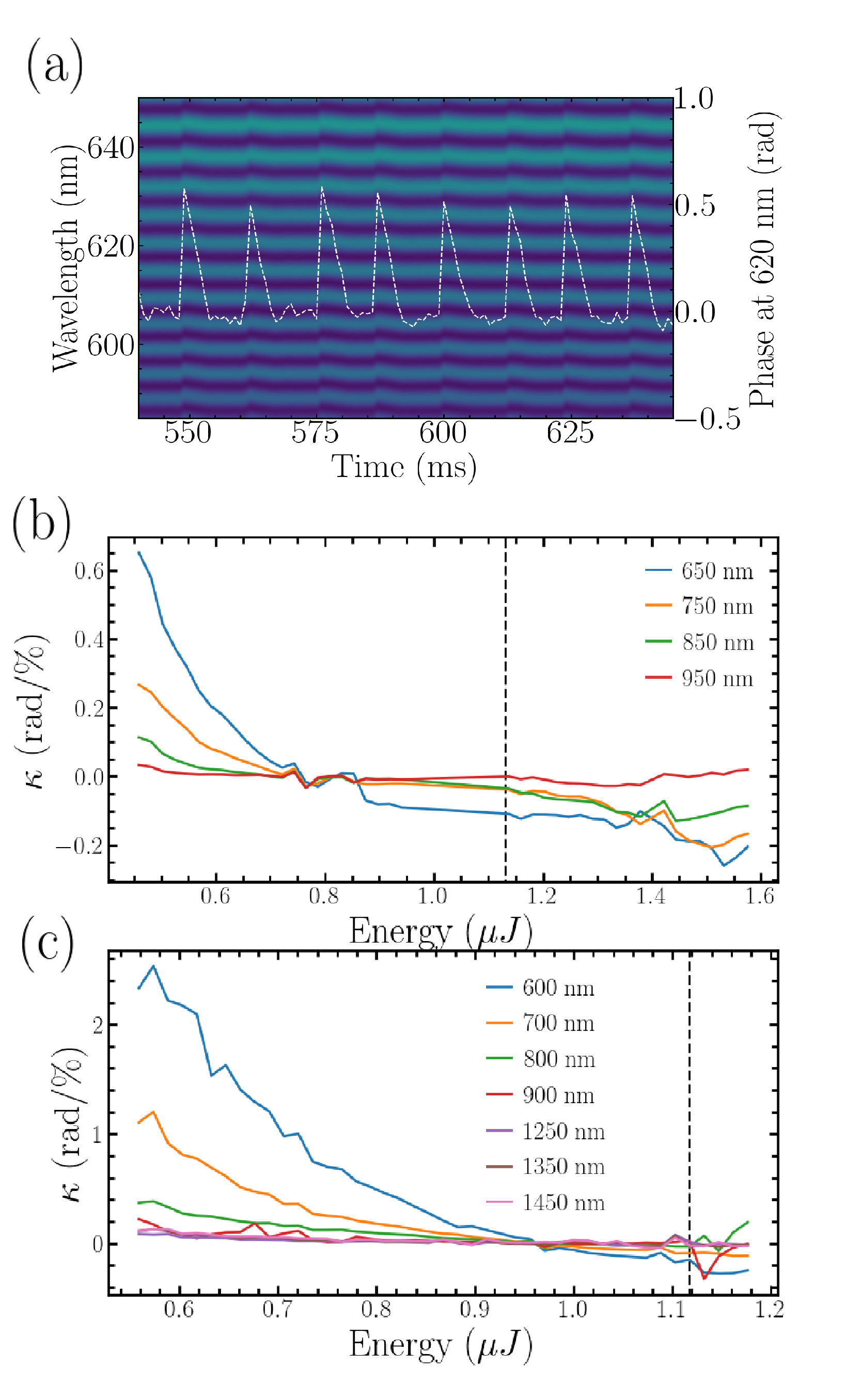}
	\caption{Typical spectrogram (a), the white dashed line is the retrieved phase at 620 nm. Amplitude-to-phase coupling coefficient for experimental configurations II (b) and III (c). The dashed black line refers to the onset of double-filamentation.}
	\label{intensityphasecoupling}
\end{figure}

Finally, the impact of the energy overshoots (Figure \ref{diagram}), that were first filtered out, are analyzed. The intensity-to-phase coupling is actually visible in the spectrogram, as wavelength-dependent phase jumps, as shown in Fig. \ref{intensityphasecoupling}a. Quantifying the intensity-to-phase coupling as a function of wavelength is of interest for two reasons. First, it plays an important role in CEP-stable sources \cite{Natile2021}. Second, a stationary point can be identified.  
We therefore extract the phase jumps values, related to the 1\% energy laser fluctuations for all experimental configurations, and compute for these laser pulses the spectrally-resolved energy-to-phase transfer coefficient ($\kappa(\lambda)$).
The results for experimental configuration II and III are shown in Figures \ref{intensityphasecoupling}b and \ref{intensityphasecoupling}c, featuring $\kappa(\lambda)$ as a function of seed energy in the test arm. 
In both cases the infrared part of the continuum seems not to be affected by the intensity fluctuations, i.e. the transfer coefficient is close to 0. In the visible part of the spectrum $\kappa(\lambda)$ is (i) larger when the energy is low or close to the onset of double filamentation, and (ii) larger for shorter wavelengths. The maximum value of $\kappa$ is much larger for experimental configuration (III) than (II): 2 rad./\% versus 0.6 rad./\% at 650 nm. Finally, in both configurations, a stationary point is identified at 0.785$\pm$0.085 $\mu$J for configuration (II) and at 0.955$\pm$0.015 $\mu$J for configuration (III). For the latter configuration, the stationary point lies just below the double-filamentation threshold. 

Table \ref{summary} gathers some significant phase stability values measured in this study (stability range, minimum phase jitter and coupling coefficients).

To conclude, by means of a modified Bellini-Hansch interferometer, we characterized the stochastic fluctuations of the spectral phase of a continuum generated via the filamentation of $\simeq$180\,fs pulses at $\simeq$1035\,nm in YAG crystals. Lenient generation conditions have been identified and it is found that even if the crystal position modifies the spectral shape, the spectral phase of the continuum remains stable across the full bandwidth of the continuum for large range of positions.
We have also quantified the chromatic dependency of the intensity-to-phase coupling coefficient. For all the studied configurations, we could identify an experimental configuration for which this coupling coefficient vanishes. In the context of CEP-stable sources, our results indicate that intrapulse coherence may contribute to the measured CEP noise, at the DFG level and/or in the f-to-2f, to a level of a few tens of mrad. Beyond CEP stability, our results may also be of interest for pulse synthesis as well as for broadened frequency combs. These results could, however, depend on the characteristics of the driving laser, and, in particular, on beam quality, temporal contrast and spectro-temporal stability. 

%\section*{Backmatter}

\section*{Funding} We acknowledge financial support from Horizon 2020 program under the Marie Skłodowska-Curie project Smart-X (GA860553), the Agence Nationale de la Recherche France (Grant ANR-19-CE30-0006-01  UNLOC), and the European Regional Development Fund (OPTIMAL).
	
\section*{Acknowledgments} The authors thank G. Steinmeyer for fruitful discussion.
	
\section*{Disclosures} The authors declare no conflicts of interest.
	
\section*{Data Availability Statement} Data underlying the results presented in this paper are not publicly available at this time but may be obtained from the authors upon reasonable request.

%\section*{References}

% Bibliography
%\bibliography{arxiv}

% Full bibliography added automatically for Optics Letters submissions; the following line will simply be ignored if submitting to other journals.
% Note that this extra page will not count against page length
%\bibliographyfullrefs{arxiv}

%Manual citation list

\end{document}